\documentclass[epjCONF]{svjour}
\usepackage{graphics}
\usepackage[latin1]{inputenc}
\newcommand{\be}{\begin{equation}}
\newcommand{\ee}[1]{\label{#1} \end{equation}}
\newcommand{\ba}{\begin{eqnarray}}
\newcommand{\ea}[1]{\label{#1} \end{eqnarray}}
\newcommand{\nl}{\nonumber \\}
\newcommand{\pd}[2]{ \frac{\partial {#1}}{\partial {#2}}  }
\newcommand{\re}[1]{(\ref{#1})}

\session-title{Hot and Cold Baryonic Matter -- HCBM 2010}
\usepackage{amsmath}
\usepackage{pifont}

\begin{document}

\title{Kinetic equilibrium and relativistic thermodynamics}
\author{P. Ván\inst{1,2}
\fnmsep\thanks{\email{vpet@rmki.kfki.hu}}}
\institute{Dept. of Theoretical Physics, KFKI Research Institute for Particle and Nuclear Physics, \\ 
        H-1525 Budapest, P.O.Box 49, Hungary; 
\and  {Dept. of Energy Engineering, Budapest Univ. of Technology and Economics},\\
  H-1111, Budapest, Bertalan Lajos u. 4-6,  Hungary}
\abstract{
Relativistic thermodynamics is treated from the point of view of kinetic theory. It is shown that the generalized Jüttner distribution suggested in \cite{BirVan10a} is compatible with kinetic equilibrium. The requirement of compatibility of  kinetic and thermodynamic equilibrium reveals several generalizations  of the Gibbs relation where  the velocity field is an independent thermodynamic variable.} 
%
\maketitle 
\section{Introduction}
\label{intro}
Relativistic thermodynamics is  the less established a-mong relativistic theories. The famous "imbroglio" of the temperature of moving bodies \cite{IsrSte80a}  indicate, that basic thermodynamic concepts may have contradictory explanation in relativity theory. The first generalization of thermodynamics in a relativistic framework was the proposal of Planck and Einstein \cite{Pla07a,Ein07a}, and were   questioned by several critical approaches more than half century later \cite{Ott63a,Bla47a,Lan66a,Lan67a}. These suggestions seemingly exclude each other, but their arguments are convincing enough to provoke a longstanding discussion that continues even today 
\cite{Req08m,Yue70a,Tre77a,Lan81a,LanMat96a,LanMat04a,CubAta07a,DunHan09a,Zyg10a,Nak09a,Nak10a}. On the other hand the experimental measurement of temperature of fast moving bodies  is an everyday practice in astrophysics and cosmology \cite{Dem85b,PeeWil68a} and in heavy ion physics \cite{Cse94b,Bir11b} and one can deal with thermal aspects  seemingly without apparent problems. From mentioned investigations it is clear that the problem requires more than a covariant form of the  physical quantities and the Gibbs relation, or a  definition of  thermodynamic bodies (moving boundaries, etc.). The interpretation of thermodynamic equilibrium among interacting bodies seems to be the most important aspect. 

The history of temperature of moving bodies shows,  that the problem  cannot be analysed by argumentation based exclusively on thermodynamics and special relativity. Moreover, statistical mechanics alone does not give a definite answer, statistical explanations were given to all of the previous different views providing different interpretations  of the macroscopic quantities \cite{CubAta07a,Lan67a}.
The reason is that the problematic point is somehow outside the usual realm of equilibrium theories. The key aspect is the interaction of moving bodies, the thermodynamics of the motion. Therefore  the problem is best investigated in the framework of  classical field theories  -  like continuum (hydrodynamics) or  kinetic theories -  where this aspect is the part of the theory and equilibrium and thermodynamic equilibrium is motion related. However, one should be aware, that  both in hydrodynamics or in kinetic theory  the definition of equilibrium is a starting point where several basic physical quantities are interpreted, and thermodynamics and thermodynamic equilibrium plays a key role. 

In our previous research we investigated stability and causality in relativistic hydrodynamics. A detailed analysis of the Second Law revealed \cite{Van08a} that assuming acceleration independent entropy production the local rest frame entropy density may depend on the absolute value of the energy momentum vector \(s(\sqrt{E^aE_a},n)\), where \(E^a=u_bT^{ab} \), with \(u^a\) is the four velocity of the fluid and \(T^{ab}\) is the energy-momentum density tensor.  Then the   consequent Gibbs relation can be written as  
\be
de+\frac{q_a}{e}dq^a = T ds+ \mu dn.
\ee{Gr_fluid}

Here \(e=u_aT^{ab}u_b\) is the local rest frame energy density, and \(q_a= \Delta_{ab}E^{b} \) is the momentum density. In the usual treatment the second term of the left hand side is missing, the entropy depends only on  the energy density and particle number density \(s(e,n)\). Our generalization eliminated the generic instability of the original Eckart theory \cite{VanBir08a} and we have proved that the stability is independent of any flow frames like Eckart or Landau-Lifshitz.  The conditions are the nonnegativity of the transport coefficients and thermodynamic stability, the  concavity of the entropy density. This is in strong contrast to the usual theory of Eckart, which is unstable \cite{HisLin85a}, or to the Israel-Stewart theory, where there are several complicated and counterintuitive conditions  beyond the above mentioned natural requirements \cite{HisLin83a,HisLin88a}. Other different recent investigations do not give similar stability properties  
\cite{GarEta09a,TsuKun08a,DenAta08a1}.

Motivated by the previous results, and with the help of a generalized internal energy and modified basic thermodynamic relations, we have analysed the problem of relativistic temperature transformations from the point of view of relativistic hydrodynamics.  A homogenization of the energy-momentum balance shows that the entropy can depend directly on the energy-momentum vector of the body \(S(E^{a},V)\) and results in the following form of the  Gibbs-relation:
\be
g_a dE^a = TdS-pdV,
\ee{Gr_hom}
where $S$ and $V$ are the entropy and volume of the body and $g_a$ is the four-vector intensive quantity conjugated to the total energy-momentum four vector. Here we did not consider changes in particle number and assumed a constant velocity. The thermodynamic body was interpreted as a homogeneous continuum, where the local rest frame is synchronized and we required also a constant average velocity. The above relation implies that the entropy   may depend on the energy momentum vector and not only on its absolute value.  We have deduced that $g^a$ is timelike and not necessarily parallel to the velocity of the body.  The appearance of a perpendicular component  resulted in  nontrivial conditions for  thermodynamic equilibrium of two thermodynamic bodies. In case of a single body one may sit into a preferred frame, determined by the energy-momentum vector, but for interacting bodies we should consider the energy-momentum of each, therefore we do not have this possibility. The condition of equilibrium requires the equality of the temperatures and the  composite velocities, determined by the vectorial intensive quantity. These velocities are not identical to the velocities of the respective bodies \cite{BirVan10a}. With this result we could give a physical background of the historical temperature transformation formulas of Einstein-Planck, Blanusa-Ott  and Landsberg. Moreover, the experimentally observed Doppler-like transformation of the spectral temperatures was derived, too \cite{CosMat95a,LanMat96a,LanMat04a}.

 The point of view of a related, but more general theory was essential.   On the other hand the second form of the Gibbs relation \re{Gr_hom} is more general than the first one \re{Gr_fluid}: the entropy depends on the 
energy-momentum vector, not only on its absolute value. The-refore in this paper we look for an independent, but related theory, and investigate whether kinetic theory is compatible with the generalized thermodynamic relations \re{Gr_fluid} and/or \re{Gr_hom}. In these researches we focus on the following aspects of the basic problems: What is the meaning of the rest frame in the form of the thermodynamic relations? How can one distinguish dissipative and nondissipative parts of physical quantities? These questions appear as the choice of flow-frames (like Eckart or Landau-Lifshitz) in relativistic hydrodynamics, arise as the existence of Brenner diffusion velocity in 
non-relativistic hydrodynamics \cite{Bre05a}, are at the root of the temperature transformation paradox and in its most general form they are related to the very meaning of covariance and the existence of absolute objects in relativistic theories \cite{Fri83b,And67b}.

\section{Basic hydrodynamics and the separation of dissipative and non-dissipative}
\label{sec:1}

In the following the  Lorentz form  is $diag(1,-1,-1,-1)$ and the units are chosen  that speed of the light and the Boltzmann constant are one $c=1$, $k_B=1$. 
For one component fluids, the basic  macroscopic fields are the particle number density four vector \(N^{a}\), the entropy density four vector \(S^a\) and the energy momentum density tensor \(T^{ab}\). The particle number and the energy-momentum are conserved in the following investigations, therefore 
\ba
\partial_a N^a(x^b) &=& 0,\label{pnum_bal}\\
\partial_b T^{ab}(x^c) &=& 0.
\ea{emom_bal}

However, the entropy is not conserved, the entropy production is zero only in thermodynamic equilibrium:
\be
\partial_a S^a(x^b) \geq 0.
\ee{entr_bal}
 With the help of a velocity field \(u^a(x^b)\) the particle number and entropy density four vectors and the energy momentum density tensor can be expressed by local rest frame quantities as
\ba
N^a &=& n u^a +j^a,\label{pnum_split}\\
S^a &=& s u^a +J^a,\label{entr_split}\\
T^{ab} &=& E^{a}u^{b}+Q^{ab} \nl
 &=&eu^au^b+q^au^b+u^aq^b+P^{ab}. 
\ea{emom_split}

Here the four component physical quantities are splitted to parallel and  perpendicular components to the local velocity,  timelike and spacelike components in the local rest frame. \(n := N^au_a\) is the particle number density, \(j^a:= \Delta^{ab}N_b\) is the diffusion current, \(s := S^au_a\) is the entropy density, \(J^a:= \Delta^{ab}N_b\) is the entropy current, \(e:=u_aT^{ab}u_b\) is the energy density, \(q^a:=\Delta^a_cT^{cb}u_b\) is the momentum density (and the energy current) and \(P^{ab}:=\Delta^a_c\Delta^b_dT^{cd}\) is the pressure tensor. Here we have introduced the u-orthogonal projection \(\Delta^{ab} := \delta^{ab}-u^au^b\). The energy-momentum tensor is symmetric and therefore the pressure tensor is also symmetric and the momentum density and the energy current are equal. Moreover, in addition
to these usual local rest frame quantities we have introduced the energy-momentum vector $E^a:=u_bT^{ab}$ and the energy-momentum current density \(Q^{ab}:=\Delta_c^aT^{cb}\) in order to clearly distinguish between momentum density $q^a:=\Delta^a_bE^{b}$ and energy current $q^a:=u_bQ^{ab}$. The forms \re{pnum_split}-\re{emom_split}  conveniently express the particle number density four vector and the energy-momentum density tensor with the help of local rest frame quantities relative to a velocity field. 

The balances \re{pnum_bal}-\re{emom_bal} can be written as
\ba
\partial_a N^a &=& \dot n + n\partial_au^a + \partial_aj^a,\label{pnumc_bal}\\
\partial_b T^{ab} &=& \dot E^a + E^a \partial_bu^b + \partial_b Q^{ab} \nl
        &=& \dot e u^a +eu^a\partial_bu^b +\dot q^a +q^a\partial_b u^b +
             e\dot u^a + \nl
        &&     u^a\partial_b q^b+q^b\partial_b u^a +\partial_bP^{ab}, 
\ea{emomc_bal}
where the dot denotes the derivative along the world lines of the velocity field (substantial time derivative) $\frac{d}{dt}:= u^a\partial_a$. 

It is considered as an evident fact that we can additively decompose the particle number density four vector and the energy-momentum tensor into a non-dissipative and a dissipative part. We usually assume, that these parts are easily distinguished in the local rest frame, because there the non-dissipative particle number density four vector \(N^a_0\) is parallel to the velocity and the non-dissipative energy momentum tensor \(T_0^{ab}\) is diagonal:
\ba
N^a_0 &=& n_0 u^a, \label{eqpnum_split}\\
T_0^{ab} &=& e_0u^au^b-p_0\Delta^{ab}.
\ea{eqmom_split}

Here $p_0$ is the static, scalar pressure, determined by the static equation of state of the fluid. Therefore the diffusion current density $j^a$  is the dissipative part of the particle number density and the momentum density/energy current \(q^{a}\) and the difference of the total and the equilibrium pressure \(\Pi^{ab} = P^{ab} + p_0\Delta^{ab}\) - the viscous pressure - are the dissipative parts of the energy momentum. 

Let us observe that with this splitting the distinction of dissipative and non-dissipative parts of the physical quantities is  related and depends on the local rest frame, the velocity field of the continuum and requires a particular thermostatics to determine the static pressure.
However, the dissipation is principally defined by the entropy production, and therefore implicitly related to the background thermostatics, that is to the concept of local equilibrium.  On the other hand,  in case of dissipation the rest frame is not determined by the local thermodynamic equilibrium, neither by any  special form of the physical quantities. One can get restrictions of the possible equilibrium entropy four vector considering simply that general information \cite{MusBor08a}. Moreover, one may consider to extend the thermodynamic state space by the velocity field. 
In this case a pure continuum approach is based on the entropy inequality and  requires  some special mathematical methods (Coleman-Noll or Liu procedures \cite{Liu72a,TriAta08a}).
 
In this paper we choose an other way and consider kinetic theory as a general background. We will explore whether  the previously mentioned generalized relativistic thermostatics introduced by pure phenome-nological arguments is compatible with kinetic theory.

\section{Equilibrium in kinetic theory}
\label{sec:2}

For the sake of simplicity we consider here a single component Boltzmann gas without external fields, \newline where the momentum vector of a particle is $p^a$ and its mass is $m=\sqrt{p^ap_a}$. The Boltzmann equation for the one particle  distribution function $f(x^a,p^a)$,  is  
\be
 p^a\partial_a f = C(f).
\ee{Bo_eq}

Here $\partial_a$ denotes the partial derivative of $f$ regarding the spacetime event $x^a$ and $C(f)$ is the following collision integral:\ 
\begin{multline}
C(x,p)= 
 \frac{1}{2} \int d\omega_1 d\omega' d\omega'_1 \left(
        f'f_1'W(p',p_1'|p,p_1)-\right.\\
        \left.ff_1W(p,p_1|p',p_1')\right), 
\label{coll_int}
\end{multline}
where $d\omega = \frac{d^3p}{p^0}$ is the momentum space measure.  $W(p,p_1|p',p_1')$ denotes the transition rate from  a  particle pair  momentum $(p'^a, p'^b_1)$ before the collision, to particle pair momentum $(p^a,p_1^b)$ after the collision. 

The particle number density vector, the energy-momentum density tensor and the entropy density vector are defined as 
\ba
N^a &:=& \int d\omega p^a f, \label{pnum_def}\\
T^{ab} &:=& \int d\omega p^ap^b f, \label{emom_def}\\
S^a &:=& \int d\omega p^a f(\ln f-1). 
\ea{{entr_def}}
The balances of these fields are calculated from the Boltzmann equation (see e.g. \cite{GroAta80b}), exploiting the property that due to the microscopic conservation laws a linear combination of a constant and the momentum four vector \(\psi(x^a,p^a)= a(x^a)+b_a(x^b)p^a\) is a collision invariant, does not change in collisions:
\begin{displaymath}
\int d\omega \psi C=0.
\end{displaymath} 
As a consequence of this property the conservation of the number of 
particles \re{pnum_bal} and energy-momentum \re{emom_bal} follow.

We may calculate the entropy production, too. Assuming momentum independent force fields and unitary scattering matrices we get a local form of the H-theorem:
\begin{multline}
\partial_aS^a = -\int d\omega \ln f \,p^a\partial_af= \\
 \frac{1}{4}\int d\omega d\omega d\omega_1d\omega'_1  \left(\frac{f'f'_1}{ff_1} -
        \ln \frac{f'f'_1}{ff_1}-1\right)\geq0.
\label{kentr_bal}\end{multline}

This entropy production vanishes if and only if the distribution function satisfies the simple functional relation:
\be
        f(x^a,p'^a)f(x^a,p'^a_1) = f(x^a,p^a) f(x^a,p_1^a).
\ee{eq_cond}

We assume that the distribution function tends to a definite limit as time progresses and the system develops into an equilibrium state where the entropy attains its maximum value. This condition, together with the requirement, that the equilibrium distribution function must be a solution of the transport equation uniquely determines its form. 

In case of binary collisions the momenta satisfies the conservation of energy-momentum
\be
p^a +p_1^a = p'^a+{p}^{\prime a}_1.
\ee{lem_cond}
Moreover, the local equilibrium distribution function \(f_0\) satisfies the functional relation \re{eq_cond}. Therefore the logarithm of the equilibrium distribution function is a  collision invariant, and necessarily has the form
\be
{f}_0(x^a,p^a) = e^{\alpha-\beta_ap^a}
\ee{edistr_def}
with arbitrary space- and time-dependent parameters \(\alpha(x^b)\) and \(\beta^a(x^b). \)
Let us remark, that the equilibrium distribution function is not a solution of the transport equation without any further ado. Substituting into \re{Bo_eq} we get, that $\alpha$ and $\beta^a$ must satisfy 
\be
p^a \partial_a \alpha + p^ap^b\partial_a\beta_b=0.
\ee{teq_cond}
\textit{Global equilibrium} is defined by the requirement, that this equation must be an identity for arbitrary values of \(p^{a}\). This implies simple initial (boundary) conditions and gives  that \(\alpha\) is a constant field and \(\beta^a\) obeys the Killing equation:
\be
\alpha(x) = const, \qquad \partial_a\beta^b + \partial_b\beta^a = 0.
\ee{kineq_cond}
\textit{Local equilibrium} considers  more involved initial (bo-undary) conditions of the transport equation and the consequent macroscopic balances are used to interpret and evaluate the equilibrium fields through the equations \re{pnum_def}-\re{emom_def}.

With the help of   the equilibrium distribution function we may calculate the particle number  balance: 
\begin{multline}
 \partial_a N_0^a = \int d\omega p^a \partial_a f_0 =  \\
 =\int d\omega p^a \partial_a f_0 -
        \int d\omega p^a p^b f_0\partial_a \beta_b = \\
    = N_0^a\partial_a\alpha -T_0^{ab} \partial_a \beta_b.
\label{pnbal_eq}\end{multline}
 
In order to get thermodynamic relations one can  calculate the Boltzmann-Gibbs entropy with the equilibrium distribution \re{edistr_def}, too:    
\be
 S^a_0 =  \int d\omega p^a f_{0}(\ln f_{0}-1)=(1-\alpha)N_0^a +\beta_bT_{0}^{ab}.
\ee{Entreq}
 \re{pnbal_eq} and \re{Entreq} results in 
\be
\partial_aS_0^a + \alpha \partial_aN^a_0 - \beta_b\partial_aT^{ab} = 0.
\ee{en_eq}

The above relations indicate a  kind of Legendre transformation between the formulas \re{pnbal_eq}  and \re{en_eq}.  However,   \re{en_eq} cannot be interpreted as a generalized Gibbs-relation, we cannot  consider the divergences  as differentials without any further ado. This relation alone does not fix the the entropy as a function of energy-momentum and particle number density. The reason is that the divergences are special derivatives and therefore the tensorial order of the coefficients \(\alpha\) and \(\beta_a\) is lower than would be in case of a natural functional relation of the differentials. We will see that several different Gibbs  relations are compatible with \re{en_eq}.
On the other hand, we may recognize, that \re{en_eq} is the balance of entropy constrained by the balances of the particle number density and the energy-momentum, where \(\alpha\) and \(\beta_a\) are Lagrange multipliers.\footnote{That later interpretation is important also beyond equilibrium, where \re{en_eq} is an inequality, therefore $\alpha$ and $\beta_a$ are Lagrange-Farkas multipliers \cite{MulRug98b,Van03a}.}

The above formulas  define the relation of kinetic theory and thermodynamics. Now we want to find the most general form of this correspondence. 
\section{Thermodynamics and kinetic theory - natural frame}

According to the usual interpretation of the equilibrium fields the absolute value of the vector \(\beta_a \) is the reciprocal temperature, its direction gives the velocity field \(\hat u_{a}\) and \(\alpha\) is the chemical potential over the temperature:
\be 
\alpha = \frac{\hat \mu}{\hat T}, \qquad \beta_a=\frac{\hat u_a}{\hat T}.
\ee{pint_def} 

In this special case \re{edistr_def} is the classical equilibrium distribution of Jüttner \cite{Jut11a}, the  relativistic generalization of the non-relativistic Maxwell-Boltzmann distribution:
\be
f_0 = e^{\frac{\hat \mu-\hat u_ap^a}{\hat T}}. 
\ee{Jutd}   

The  momentum space integral defining the equilibrium particle number density exist only if  \(\beta_a\) is a timelike vector \cite{Lev96a}. Therefore  the particle number density four vector \(N^{a}_0\) in equilibrium is proportional to \(\beta^{a}\)
\be
  N_0^a = \int d\omega p^a e^{\alpha-\beta_ap^a} =  A\beta^a, 
\ee{eqnbal}
where \(A\) is a scalar field. 

This observation is the reason, why in local equilibrium one may expect zero diffusion current density \(j^{a} = \Delta^{a}_bN^b =0\). In this case it is possible to 
\textit{define} a velocity field by particle number density four vector assuming, that \(N_0^a=\hat n\hat u^a = A \beta^a\), where $\hat n$ is the equilibrium local rest frame particle number density. In this case the vector \(\beta^a\) is parallel to the (equilibrium) velocity field, and  
\be 
\hat u^a:=\frac{\beta^a }{\sqrt{\beta_b\beta^b}}, \quad 
\hat T :=\frac{ 1}{\sqrt{\beta_b\beta^b}} \quad 
\hat n:=A \sqrt{\beta_b\beta^b}
\ee{class_def} 
are consistent definitions.  We will call this choice of the velocity
field as \textit{natural equilibrium frame.} 

A similar calculation gives the equilibrium energy-momentum tensor. Together with the particle number balance they are written
as\ba
  N_0^a &=& \int d\omega p^a e^{\alpha-\beta_ap^a} =  \hat n\hat u^a, \label{eqnbalt}\\
 T_0^{ab} &=& \int d\omega p^a p^bf_0 = \hat e \hat u^a \hat u^b - \hat p\hat \Delta^{ab}.
\ea{eqemombal}

Here the diagonal form is the consequence of the special  distribution function and we obtain the usual relations \cite{Cse94b,GroAta80b}: 
\ba
\hat n &=& 4 \pi m^2\hat T K_2\left(\frac{m}{\hat T}\right) e^{\frac{\hat\mu}{\hat T}}, \label{ntraf}\\
\hat p &=& \hat n\hat T,\\
\hat e &=& 3 \hat n \hat T + m\hat n \frac{K_1\left(\frac{m}{\hat T}\right)}
    {K_2\left(\frac{m}{\hat T}\right)}.
\ea{etraf}
 
 Substituting \re{eqnbal} and \re{eqemombal} into \re{Entreq} 
we get  \be
\! S^a =\hat u^a \left ( (1\!-\!\alpha)\hat n + \frac{\hat e}{\hat T}\right)  .
\ee{sid}
 
Therefore the direction of the entropy four vector is parallel to the natural velocity field. Substituting \re{eqnbalt} and \re{eqemombal} into \re{en_eq} results in \begin{multline}
\partial_aS_0^a + \alpha \partial_aN^a_0 - \beta_b\partial_aT^{ab} = \\
 \hat u^a \left(\partial_a \hat s+ \alpha\partial_a \hat n -
    \frac{1}{\hat T}\partial_a \hat e\right)+ \\
 \partial_a\hat u^a \left(\hat s +\alpha \hat n -\frac{1}{\hat T}(\hat e+\hat p) \right)=0. \label{devbasic1}
\end{multline}

 We may require that the equality is valid for arbitrary velocity fields and  velocity gradients. Therefore the expression in parenthesis at the second line of the previous formula gives that the entropy density in the natural local rest frame is the function of the energy density and the particle number density \(\hat s = \hat s(\hat e, \hat n) \) and the  partial derivatives are:
\be
\pd{\hat s}{\hat n} = \alpha=\frac{\hat \mu}{\hat T}, \qquad
\pd{\hat s}{\hat e} = \frac{1}{\hat T}. 
\ee{classpd}
The consequent classical Gibbs relation for the differentials reads:
\be
 d\hat e = \hat Td\hat s+\hat \mu d\hat n.
\ee{Grelc}
The second line of \re{devbasic1} gives the potential relation, fixing the Legendre transformation properties and defining the pressure:
\be
 \hat p := \hat T\hat s +\hat \mu \hat n -\hat e. 
\ee{potrelc}
Hence the thermodynamic interpretation of the kinetic quantities is complete and one may get the Gibbs-Duhem relation,   by the total Legendre transformation of \re{Grelc}:
\be
  d\hat p = \hat s d\hat T + \hat n d\hat \mu. 
\ee{GDrelc}

Finally we should recognize, that according to  \re{sid}, the corresponding material is an ideal gas, providing the relation \(\hat p = \hat T\hat n\).

One can observe, that in the natural frame interpretation we have defined a convenient four velocity field and hence a rest frame of the material and all the thermodynamic quantities and relations are defined relative to that.

\section{Thermodynamics and kinetic theory - general frame}

We may assume, that the velocity field \(u^{a}\) of an ideal fluid is not parallel to  \(\beta_a\).
We will call this choice of the velocity
field as \textit{general equilibrium frame. }In this case  \(\beta_a\) can be decomposed into parallel and perpendicular components:\be
\beta^a = \frac{u^a +w^a}{T},
\ee{decbeta}
where we have used the notation\ \(\frac{1}{T}:= \beta_au^a\) and \(w^a := \frac{\Delta^a_b \beta^b}{\beta_cu^c}\). It is convenient to introduce also \(g^a = u^a+w^a\). We can see that 
$g^a u_a = 1$  and \(w^{a}u_a=0\), moreover $w_aw^a<1$, because \(\beta^a\) and therefore \(g^a\) are timelike.
Now the  equilibrium distribution can be written as   
\be
f_0 = e^{\frac{\mu-p^a(u_a+w_a)}{T}}. 
\ee{genJut_def}
 
Then we can calculate the equilibrium particle number four vector and also the energy momentum tensor, recognizing, that the relations of the different quantities
in natural and general frames are 
\be
T=\sqrt{1-w^2} \hat T, \quad \mu=\sqrt{1-w^2} \hat \mu,
 \quad \hat u^a =\frac{u^a+w^a}{\sqrt{1-w^2}},
\ee{trafo1}
where \(w^2= -w^aw_a\).

Then the equilibrium particle number four vector $N^a_0$ and energy-momentum tensor \(T_0^{ab}\)  can be written in the general frame as
\ba
N_0^a &=& ng^{a} = n u^{a} + n w^a\ , \label{ndpnumg}\\
T_0^{ab} &= & (e+p)g^ag^b - p \delta^{ab} \nonumber\\
    &=& e u^{a} u^b + q^{a}u^b+q^bu^a - p\Delta^{ab}+\frac{q^aq^b}{e+p}.
\ea{ndemomg}
where
\ba
 n &=&\frac{\hat n}{\sqrt{1-w^2}},\\
 p &=& \hat p, \\
 e &=& \frac{\hat e+\hat p w^2}{1-w^2}, \\
 q^a &=& (e+p)w^a, \label{idheat}\\ 
 E^a &=& eu^a + (e+p)w^a.
\ea
 TThe energy and particle number densities and the pressure can be calculated from the corresponding forms in the natural frame.  Substituting \re{ndpnumg} and \re{ndemomg} into \re{Entreq} 
we get  
\be
\! S^a =\frac{u^a+w^a}{T} \left ( nT -\mu n + E^ag_a\right)  ,
\ee{sid2}
where we have introduced $\mu:=\sqrt{1-w^2}\hat \mu$. The entropy current density in the general frame is parallel to \(g^{a}\).  The thermodynamic relations are calculated by substituting \re{ntraf} and \re{etraf} into \re{en_eq}, initially assuming only that \(S^{a} = sg^a\):
\begin{multline}
\partial_aS_0^a + \alpha \partial_aN^a_0 - \beta_b\partial_aT^{ab} = \\
 \frac{g^a}{T}  \left(T\partial_a  s+ \mu\partial_a n - 
    g_b \partial_a E^b - pw^b \partial_au_b\right)+ \\
 \frac{\partial_a g^a}{T} \left(T s +\mu n -g_bE^b- p \right)=0. \label{devbasic2}
\end{multline}

The second line of the previous expression shows, that the entropy density in the natural local rest frame is the function of the particle number density, the energy-momentum density four vector and the velocity  \(s =  s(E^a,  n, u^a).  \) The related partial derivatives are:
\be
\pd{s}{n} = \alpha=\frac{\mu}{T}, \quad
\pd{s}{E^a} = \frac{g_a}{T}, \quad
\pd{s}{u^a} = \frac{pw_a}{T}.
\ee{classpd1}
The consequent Gibbs relation for the differentials is:
\be
 g_{a}d E^{a} = Tds + \mu d n -p w_{a} du^{a}.
\ee{Gr_res}
According to this relations,  the change of the velocity directly contributes to the thermodynamic quantities. 

The expression in the parenthesis of the third line of \re{devbasic2} is the potential relation, fixing the Legendre transformation properties  as:
\be
  p =  Ts +\mu  n -g_aE^a. 
\ee{potrelg}
Then the Gibbs-Duhem relation follows:
\be
dp + E^{a} dg_{a}= sdT + nd\mu -p u_{a} dw^{a}.
\ee{GDrelg}

Finally \re{sid2} shows, that the corresponding material is an ideal gas, providing the relation \( p =  T n\). This equation of state does not change, is the same both in the natural and general frames. 

One can check the above expressions directly, recognizing, that
\ba
s(E^a, n)&=&\frac{1}{g}\hat s(ng,E^ag_a), \\
p(g^a,T,\mu)&=&\hat p\left(\frac{T}{g},\frac{\mu}{g}\right).
\ea{ }
Here \(g=\sqrt{1-w^2}=\sqrt{g_ag^a}\). However, these relations alone, without the constrained entropy balance \re{en_eq}, are not enough to fix a unique form of the differentials.  

One may see more  the role of the pressure and the velocity change - the acceleration - related term in the Gibbs relation \re{Gr_res}, if  the energy-momentum density vector is written in  the form \(E^a = eu^a + q^a\). In this case we get
\begin{multline}
 Tds + \mu d n = g_{a}d E^{a} + p w_{a} du^{a} =\\
  (u_a +w_a)d(eu^a +q^a) + p w_{a} du^{a} =\\
  de+w_a dq^a +((e+p)w_a -q_a)du^a.  
  \label{traf}\end{multline}
Therefore taking into account \re{idheat} we can free ourselves from the 
direct pressure and velocity dependence and get 
\be
de +w_adq^a = Tds +\mu dn.
\ee{Grelsimp}

In this relation the velocity field dependence is not apparent.

\section{Summary and conclusions}

It is well known, that thermodynamic relations are fixed to the material and therefore they are best expressed by rest frame quantities. This is a hidden aspect in nonrelativistic hydrodynamics \cite{Mat86a}, but an apparent property of relativistic theories. In thermodynamic equilibrium there is a single, distinguished velocity field, and a corresponding rest frame,  where all motion related physical quantities are simple.   

In this paper we have investigated whether a novel concept of relativistic thermodynamics based on a generalized Gibbs relation is compatible with kinetic theory. Kinetic theory distinguishes a\emph{ natural frame},  a  velocity field in equilibrium, fixed by the direction of \(\beta_a \). The entropy and particle number density four vectors are parallel to this velocity and the energy-momentum density generated accordingly. Therefore this natural frame corresponds both to  Eckart and Landau-Lifshitz frames of dissipative fluids \cite{Eck40a3,LanLif59b}. In a natural frame the thermodynamic relations are very similar to the nonrelativistic ones. 

However, one may introduce a \textit{general frame}, a velocity field independently of \(\beta_a\), and  investigate the consequent thermodynamic relations. The calculations are  transformations of the equations calculated in the natural frame, but the resulted  thermodynamic relations contain additional terms. In particular the energy-momentum vector appears as a natural variable - similarly to \re{Gr_hom} - and the velocity field plays an explicit role in the relations.  

Therefore we have shown that kinetic theory is compatible with the generalized Gibbs relation suggested to resolve stability problems of hydrodynamics and paradoxes related to the temperature of moving bodies.  The  direct velocity dependence of the Gibbs relation \re{Gr_res} in the general frame generalizes our previous results, where we either assumed that the entropy production is independent of acceleration \cite{Van08a} or considered a  thermodynamic body moving with a uniform velocity \cite{BirVan10a}. 

If \re{Gr_res} (and \re{potrelg})  are introduced in a hydrodynamic theory, the concept of local equilibrium is modified. Because the constitutive relations of dissipative fluids (viscosity, heat conduction, etc..) express the tendency toward local equilibrium,  the  entropy production and thermodynamic fluxes and forces can be modified, too.  \section{Acknowledgement}   
The work was supported by the grant Otka K81161. The author thanks T. S. Biró, E. Molnár, W. Muschik and H.H.~v. Borzeszkowski for the valuable discussions.

\bibliographystyle{unsrt}

\begin{thebibliography}{10}

\bibitem{BirVan10a}
T.~S. B\'ir\'o and P.~V\'an.
\newblock About the temperature of moving bodies.
\newblock {\em EPL}, 89:30001, 2010.

\bibitem{IsrSte80a}
W.~Israel and J.~M. Stewart.
\newblock Progress in relativistic thermodynamics and electrodynamics of
  continuous media.
\newblock In A.~Helde, editor, {\em General relativity and gravitation (One
  hundred years after the birth of {A}lbert {E}instein)}, volume~2, chapter~13,
  pages 491--525. Plenum Press, New York and London, 1980.

\bibitem{Pla07a}
M.~Planck.
\newblock Zur {D}ynamik bewegter {S}ysteme.
\newblock {\em Sitzungsberichten der k\"onigliche Preussen Akademie der
  Wissenschaften}, pages 542--570, 1907.

\bibitem{Ein07a}
A.~Einstein.
\newblock {\"U}ber das {R}elativit\"atsprinzip und die aus demselben gezogenen
  {F}olgerungen.
\newblock {\em Jahrbuch der Radioaktivit\"at und Elektronik}, 4:411--462, 1907.

\bibitem{Ott63a}
H.~Ott.
\newblock Lorentz-{T}ransformation der {W}\"arme und der {T}emperatur.
\newblock {\em Zeitschrift f\"ur Physik}, 175:70--104, 1963.

\bibitem{Bla47a}
D.~Blanu\v{s}a.
\newblock Sur les paradoxes de la notion d'\'energie.
\newblock {\em Glasnik mat. fiz.; astr.}, 2(4-5):249--50, 1947.

\bibitem{Lan66a}
P.~Landsberg.
\newblock Does a moving body appears cool?
\newblock {\em Nature}, 212:571--573, 1966.

\bibitem{Lan67a}
P.~Landsberg.
\newblock Does a moving body appears cool?
\newblock {\em Nature}, 214:903--4, 1967.

\bibitem{Req08m}
M.~Requardt.
\newblock Thermodynamics meets special relativity - or what is real in physics?
\newblock 2008.
\newblock arXiv:0801.2639v1[gr-qc].

\bibitem{Yue70a}
C.~K. Yuen.
\newblock Lorentz transformation of thermodynamic quantities.
\newblock {\em American Journal of Physics}, 38(2):246--252, 1970.

\bibitem{Tre77a}
Von H.-J. Treder.
\newblock Die {S}trahlungs-{T}emperatur bewegter {K}\"orper.
\newblock {\em Annalen der Physik}, 7(34/1):23--29, 1977.

\bibitem{Lan81a}
P.~T. Landsberg.
\newblock Einstein and statistical thermodynamics {I}: relativistic
  thermodynamics.
\newblock {\em Annals of Physics}, 56:299--318, 1970.

\bibitem{LanMat96a}
P.~Landsberg and G.~E.~A. Matsas.
\newblock Laying the ghost of the relativistic temperature transformation.
\newblock {\em Physics Letters A}, 223:401--403, 1996.

\bibitem{LanMat04a}
P.~Landsberg and G.~E.~A. Matsas.
\newblock The impossibility of a universal relativistic temperature
  transformation.
\newblock {\em Physica A}, 340:92--94, 2004.

\bibitem{CubAta07a}
D.~Cubero, J.~Casado-Pascual, J.~Dunkel, P.~Talkner, and P.~H\"anggi.
\newblock Thermal equilibrium and statistical thermometers in special
  relativity.
\newblock {\em Physical Review Letters}, 99:170601, 2007.

\bibitem{DunHan09a}
J.~Dunkel and P.~H\"anggi.
\newblock Relativistic {B}rownian motion.
\newblock {\em Physics Reports}, 471(1):1--73, 2009.
\newblock arXiv:0812.1996v2.

\bibitem{Zyg10a}
R.~Zygadlo.
\newblock Thermodynamical quantities and relativity.
\newblock {\em Acta Physica Polonica}, 41(5):1073--1081, 2010.

\bibitem{Nak09a}
T.~K. Nakamura.
\newblock Lorentz transform of black-body radiation temperature.
\newblock {\em EPL}, 88:20004, 2009.

\bibitem{Nak10a}
T.~K. Nakamura.
\newblock Thermodynamics of extended bodies in special relativity.
\newblock {\em EPL}, 89:40007, 2010.

\bibitem{Dem85b}
Marek Demia\'~nski.
\newblock {\em Relativistic astrophysics}.
\newblock PWN-Polish Scientific Publishers - Pergamon Press, Warszawa - Oxford,
  etc., 1985.

\bibitem{PeeWil68a}
P.~J.~E. Peebles and D.~T. Wilkinson.
\newblock Comment on the anisotropy of the primeval fireball.
\newblock {\em Physical Review}, 174(5):2168, 1968.

\bibitem{Cse94b}
Csernai L.
\newblock {\em Introduction to relativistic heavy ion physics}.
\newblock John Wiley and Sons, Chicester-etc., 1994.

\bibitem{Bir11b}
T.~S. Bir\'o.
\newblock {\em Is there a temperature?}
\newblock Springer, 2011.

\bibitem{Van08a}
P.~V\'an.
\newblock Internal energy in dissipative relativistic fluids.
\newblock {\em Journal of Mechanics of Materials and Structures},
  3(6):1161--1169, 2008.

\bibitem{VanBir08a}
P.~V\'an and T.~S. B\'\i{}r\'o.
\newblock Relativistic hydrodynamics - causality and stability.
\newblock {\em The European Physical Journal - Special Topics}, 155:201--212,
  2008.

\bibitem{HisLin85a}
W.~A. Hiscock and L.~Lindblom.
\newblock Generic instabilities in first-order dissipative relativistic fluid
  theories.
\newblock {\em Physical Review D}, 31(4):725--733, 1985.

\bibitem{HisLin83a}
W.~A. Hiscock and L.~Lindblom.
\newblock Stability and causality in dissipative relativistic fluids.
\newblock {\em Annals of Physics}, 151:466--496, 1983.

\bibitem{HisLin88a}
W.~A. Hiscock and L.~Lindblom.
\newblock Stability in dissipative relativistic fluid theories.
\newblock {\em Contemporary Mathematics}, 71:181--220, 1988.

\bibitem{GarEta09a}
A.~L. Garcia-Perciante, L.~S. Garcia-Colin, and A.~Sandoval-Villalbazo.
\newblock On the nature of the so-called generic instabilities in dissipative
  relativistic hydrodynamics.
\newblock {\em General Relativity and Gravitation}, 41(7):1645--1654, 2009.

\bibitem{TsuKun08a}
K.~Tsumura and T.~Kunihiro.
\newblock Stable first-order particle-frame relativistic hydrodynamics for
  dissipative systems.
\newblock {\em Physics Letters B}, 668(5):425--428, 2008.

\bibitem{DenAta08a1}
G.~S. Denicol, T.~Kodama, T.~Koide, and Ph. Mota.
\newblock Stability and causality in relativistic dissipative hydrodynamics.
\newblock {\em Journal of Physics G - Nuclear and Particle Physics},
  35(11):115102, 2008.

\bibitem{CosMat95a}
S.~S. Costa and G.~E.~A. Matsas.
\newblock Temperature and relativity.
\newblock {\em Physics Letters A}, 209:155--159, 1995.

\bibitem{Bre05a}
H.~Brenner.
\newblock Kinematics of volume transport.
\newblock {\em Physica A}, 349:11--59, 2005.

\bibitem{Fri83b}
M.~Friedman.
\newblock {\em Foundations of Space-Time Theories (Relativistic Physics and
  Philosophy of Science)}.
\newblock Princeton University Press, Princeton, New Jersey, 1983.

\bibitem{And67b}
J.~L. Anderson.
\newblock {\em Principles of relativity physics}.
\newblock Academic Press, 1967.

\bibitem{MusBor08a}
W.~Muschik and H.H.~v. Borzeszkowski.
\newblock Entropy identity and equilibrium conditions in relativistic
  thermodynamic.
\newblock {\em General Relativity and Gravitation}, 41(6):1285--1304, 2008.

\bibitem{Liu72a}
I-Shih Liu.
\newblock Method of {L}agrange multipliers for exploitation of the entropy
  principle.
\newblock {\em Archive of Rational Mechanics and Analysis}, 46:131--148, 1972.

\bibitem{TriAta08a}
V.~Triani, C.~Papenfuss, V.~A. Cimmelli, and W.~Muschik.
\newblock Exploitation of the {S}econd {L}aw: {C}oleman-{N}oll and {L}iu
  procedure in comparison.
\newblock {\em Journal of Non-Equilibrium Thermodynamics}, 33:47--60, 2008.

\bibitem{GroAta80b}
W.A. De~Groot, van Leeuwen, and Ch.~G. van Weert.
\newblock {\em Relativistic Kinetic Theory}.
\newblock North Holland, Amsterdam, 1980.

\bibitem{MulRug98b}
I.~M\"uller and T.~Ruggeri.
\newblock {\em Rational Extended Thermodynamics}, volume~37 of {\em Springer
  Tracts in Natural Philosophy}.
\newblock Springer Verlag, New York-etc., 2nd edition, 1998.

\bibitem{Van03a}
P.~V\'an.
\newblock Weakly nonlocal irreversible thermodynamics.
\newblock {\em Annalen der Physik (Leipzig)}, 12(3):146--173, 2003.
\newblock (cond-mat/0112214).

\bibitem{Jut11a}
F.~J\"uttner.
\newblock Das maxwellsche gesetz der geschwindigkeitsverteilung in der
  relativtheorie.
\newblock {\em Annalen der Physik}, 339(5):856--882, 1911.

\bibitem{Lev96a}
C.~D. Levermore.
\newblock Moment closure hierarchies for kinetic theories.
\newblock {\em Journal of Statisticsl Physics}, 83(5/6):1021--1065, 1996.

\bibitem{Mat86a}
T.~Matolcsi.
\newblock On material frame-indifference.
\newblock {\em Archive for Rational Mechanics and Analysis}, 91(2):99--118,
  1986.

\bibitem{Eck40a3}
Carl Eckart.
\newblock The thermodynamics of irreversible processes, {III}. {R}elativistic
  theory of the simple fluid.
\newblock {\em Physical Review}, 58:919--924, 1940.

\bibitem{LanLif59b}
L.~D. Landau and E.~M. Lifshitz.
\newblock {\em Fluid mechanics}.
\newblock Pergamon Press, London, 1959.
\end{thebibliography}

\end{document}